\documentclass[conference]{IEEEtran}
\IEEEoverridecommandlockouts

\usepackage{cite}
\usepackage{amsmath,amssymb,amsfonts}
\usepackage{algorithmic}
\usepackage{graphicx}
\usepackage{textcomp}
\usepackage{xcolor}

\usepackage{url}
\usepackage{adjustbox}
\usepackage{verbatim}
\usepackage{setspace}

\def\BibTeX{{\rm B\kern-.05em{\sc i\kern-.025em b}\kern-.08em
    T\kern-.1667em\lower.7ex\hbox{E}\kern-.125emX}}
\begin{document}

\title{Electronically Reconfigurable Pinching Antennas for Millimeter-Wave Communication in LoS and NLoS Environments}

\author{\IEEEauthorblockN{Fernando Plata-Orozco}
\IEEEauthorblockA{\textit{School of Electrical, Computer and Energy Engineering} \\
\textit{Arizona State University}\\
Tempe, AZ 85287, USA \\
fplataor@asu.edu}
\and
\IEEEauthorblockN{Mohammadreza F. Imani}
\IEEEauthorblockA{\textit{School of Electrical, Computer and Energy Engineering} \\
\textit{Arizona State University}\\
Tempe, AZ 85287, USA \\
Mohammadreza.Imani@asu.edu}
}

\maketitle

\begin{abstract}
This paper presents the design and operation of an electronically reconfigurable pinching antenna (E-pinching antenna) and examines its capability to establish controllable millimeter-wave links that can circumvent blockages. The antenna consists of a low-loss rectangular dielectric waveguide that leaks energy through modular varactor-loaded \textit{pinches} to form tunable radiation points. Full-wave simulations show that adjusting the varactor capacitances dynamically tunes the radiated power and transmission coefficient to a receiving patch antenna. We investigate a multi-user scenario by activating two radiation modules along the waveguide to serve spatially separated receiving antennas isolated by a metallic partition (blockage). Simulation results confirm the capability to establish links to both LoS and NLoS users with reduced propagation losses. The proposed architecture enables a scalable and electronically controllable distributed antenna platform for adaptive mmWave wireless networks.
\end{abstract}


\section{Introduction}
The sixth generation of telecommunications (6G) is projected to have a massive number of users with exponentially growing data demands \cite{saad2019vision}. This vision has motivated exploration of higher-frequency bands that offer larger bandwidths \cite{1220566}. Increasing the operating frequency, however, brings its own challenges, such as severe propagation losses and blockage, which require physical-layer modifications \cite{rappaport2019wireless}. To address these limitations, technologies such as Reconfigurable Intelligent Surfaces (RISs), Fluid Antenna Systems (FASs), and Dynamic Metasurface Antennas (DMAs) have been proposed to mitigate blockage effects, enhance spatial diversity, and improve spectral efficiency \cite{RIS,shlezinger2021dynamic, IMS,Fluid,CommunicationSuperhighways,trichopoulos2022design,basar2019wireless,Kiat,zheng2025reconfigurableantennas6gtechnologies}. Despite these advances, many challenges associated with millimeter-wave (mmWave) wireless communication links, especially in indoor environments, remain unresolved \cite{bai2014coverage,sulyman2016directional,kela2016location}. 

\begin{figure}[!t]
\centering
\includegraphics[width=\linewidth]{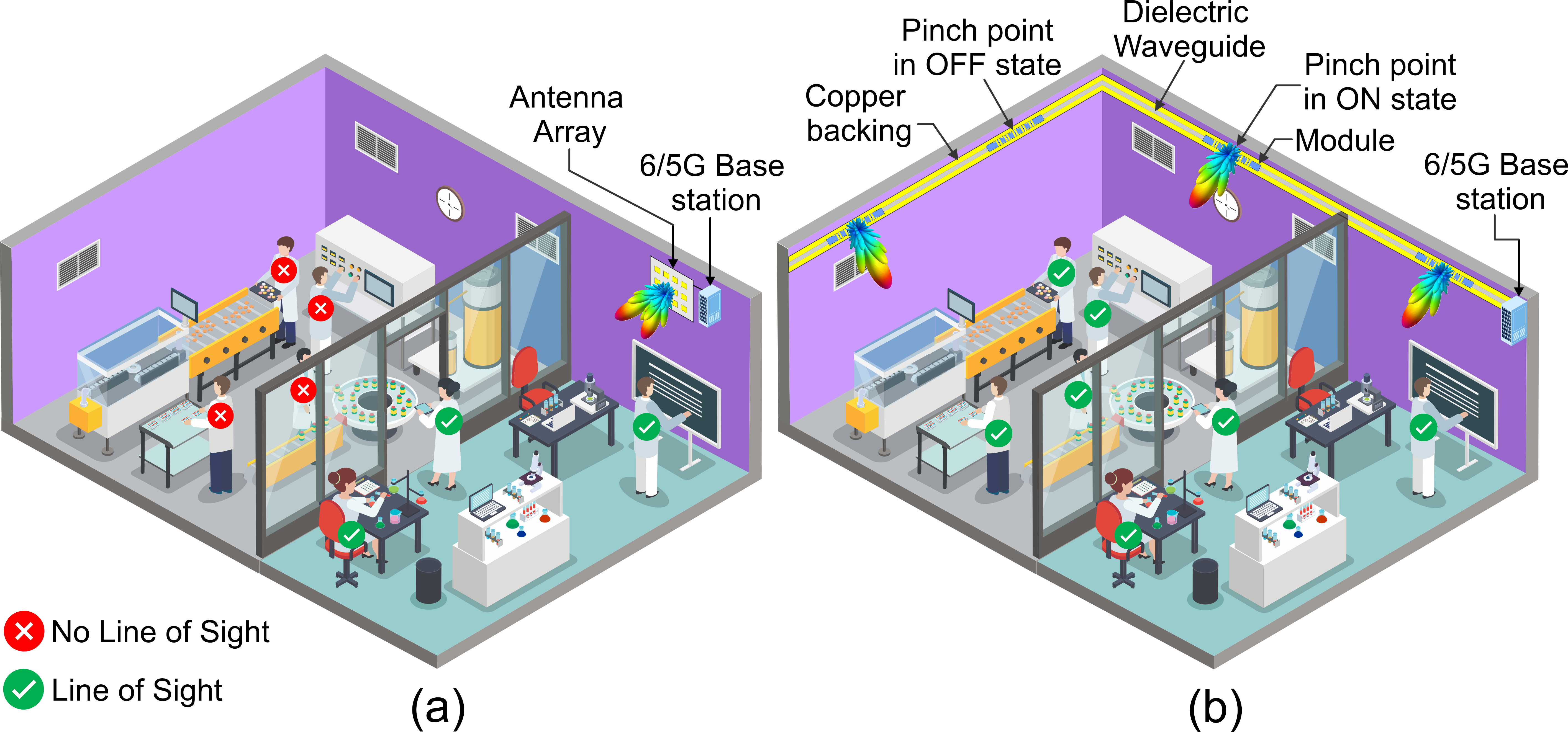}
\caption{An example of a mmWave wireless communication network using (a) conventional methods and (b) the proposed E-pinching antennas.}
\label{fig:setup}
\end{figure}

Pinching antennas are an emerging type of FAS \cite{FASPinching} that can form radiation points near intended users, enabling a line-of-sight (LoS) link with minimal propagation losses. The operating principle of a pinching antenna is based on two key factors: 1) using a dielectric waveguide to guide the electromagnetic signals near the intended users (instead of free space propagation), and 2) using a mechanical perturbation of the dielectric waveguide (e.g., by pinching it) which causes a portion of the guided signal to leak, creating a radiation point \cite{DOCOMO,Yang2025Pinching}. By changing the size of the pinching device, e.g., by using clamps with multiple independently controlled mechanical pins, the radiation intensity can be controlled \cite{Multipin}. 

The potential role of pinching antennas in mmWave wireless communication is illustrated in Fig. \ref{fig:setup} and compared with that of conventional methods. As shown in Fig. \ref{fig:setup}(a), the conventional scheme utilizes an access point with a sophisticated phased array to track users and provide them with highly directive beams. If the LoS to the users is blocked, the communication signal may be lost. Instead, a pinching antenna, as shown in Fig. \ref{fig:setup}(b), forms a low-loss data highway around the indoor environment. Near the intended users, the dielectric waveguide is pinched to form the communication link with minimal propagation loss while avoiding blockage.

While interest in pinching antennas is growing, existing implementations rely on mechanical perturbations, which are bulky, slow, and difficult to scale. To overcome this limitation, we present an electronically reconfigurable pinching antenna (E-pinching antenna) that eliminates the need for mechanical components. As an illustrative example, we design this antenna for operation at 60 GHz. Using full-wave simulations in a small-scale configuration, we demonstrate that the proposed architecture establishes controllable mmWave links with both LoS and NLoS users.

\section{Antenna Design}

The proposed electronically reconfigurable pinching antenna is based on a low-loss dielectric waveguide operating at 60 GHz. The waveguide consists of polytetrafluoroethylene (PTFE) ($\varepsilon_r = 2.1$, $\tan\delta = 0.0004$) \cite{RodWGPTFELoss} and is designed to enable low-loss propagation of the fundamental hybrid $E_{11}^{X}$ mode \cite{WGSlabAntenna}. The theory behind the design of this dielectric waveguide in the desired modes is described in \cite{WGBook, BalanisWG}. To excite the dielectric waveguide in the desired hybrid mode, the excitation port cross-section is defined as larger than the waveguide cross-section. In Ansys HFSS, this configuration allows the simulator to automatically transform the input field into the hybrid propagation mode supported by the dielectric waveguide. Since this mode-conversion technique has been extensively studied in the literature, we do not discuss its implementation details here; interested readers are referred to \cite{HE11conversion1, HE11conversion2, HE11conversion3,WGConversion4,WGConversion5}. By confining electromagnetic energy within the dielectric structure, the waveguide greatly reduces propagation losses—3.2 dB—compared to 68 dB for free-space transmission over 1 meter, making it well-suited for mmWave signal transport in indoor environments.

\begin{figure}[!t]
\centering
\includegraphics[width=3.5in]{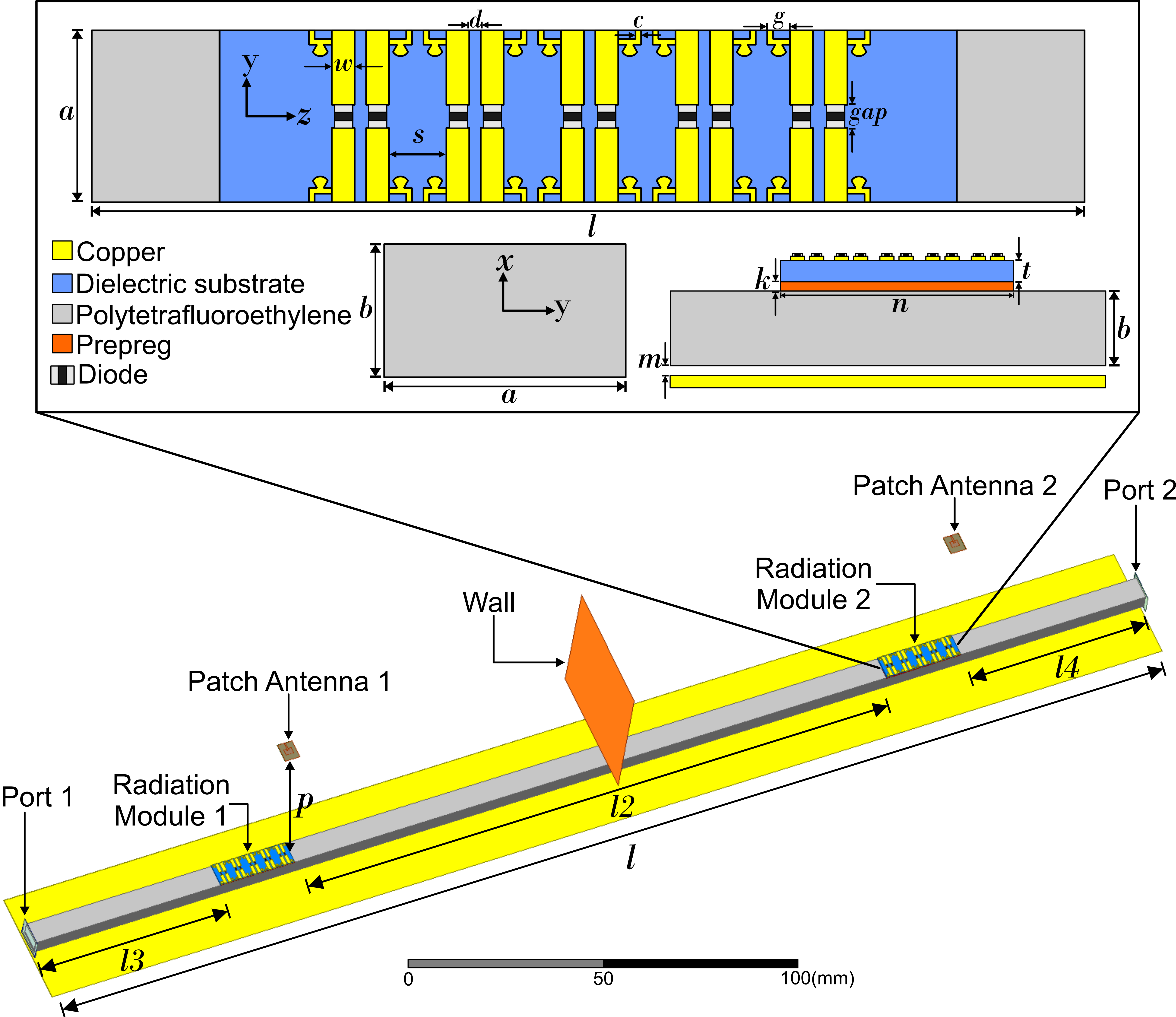}
\caption{The general configuration of the E-pinching antenna and the simulation setup. The electronically reconfigurable module is shown in the inset. A small-scale multi-user LoS and NLoS wireless communication system is also plotted.}
\label{fig_2}
\end{figure}

To enable localized radiation zones, perturbations in the form of metallic strips are positioned on top of the dielectric waveguide \cite{WGRodAntenna,WGSlabAntenna,LWAsHum}. To control radiation, we introduce a varactor diode between the strips as shown in Fig. \ref{fig_2}. By varying the varactor capacitance, we can adjust the radiated power and control the amount of power extracted from the waveguide. The proposed design employs five metallic strips. This parameter can be optimized depending on the application and is left to future work. To facilitate the fabrication and integration of these radiation points, they are assumed to be implemented on a modular printed circuit board, as shown in the inset of Fig. \ref{fig_2}. RF filtering is incorporated into the radiation circuit using radial stubs to suppress RF coupling to the DC biasing lines. When placed on top of the dielectric waveguide, the module creates a controllable radiation point at the desired location. Preliminary results on this module and its performance were reported in \cite{Epinch}. The present work extends and refines that preliminary study; specifically, we have corrected the module geometry and switchable component model. In addition, we extend the study by adding a reflector, examining wireless channels between the pinching antenna and one or two users.

In most practical scenarios, this antenna would be placed near walls, headsets, etc., which would require a unidirectional pattern. A metallic backing plane is thus introduced behind the antenna at a distance of $m=3$ mm (optimized via a parametric sweep; not shown for brevity). It is important to emphasize that this reflector cannot be placed closer to the waveguide since it would interact with the fringing fields and perturb the guided modes of the waveguide. This distance can easily be maintained using an expanded-polystyrene spacer or 3D-printed mount. 

The design parameters for the E-pinching antenna are shown in Fig. \ref{fig_2}, where the dimensions of the antenna are $a = 6$ mm, $b = 4$ mm and $l = 60$ mm. The dielectric substrate supporting the metallic strips is Rogers RT/Duroid 5880, with a thickness of $t = 0.252$ mm, relative permittivity $\varepsilon = 2.2$, and loss tangent $\delta = 0.0009$. The strip parameters are $s = 2$ mm, $d = 0.4$ mm, $w = 0.8$ mm, and $gap = 0.8$ mm. The dimensions of the DC biasing line with the radial stub filter are $c = 0.2$ mm, $g = 0.8$ mm. The module size is $n=20$ mm. To better emulate a practical setting, we also included a $k = 0.102$ mm thick prepreg layer made of Rogers RO4450 with a relative permittivity of $\varepsilon = 3.52$, and a loss tangent of $\delta = 0.004$. This layer is optional and can be eliminated depending on the application. 

The choice of the switchable component can significantly affect antenna performance. Here, we evaluate the E-pinching concept independent of a specific switching component. Since the proposed operation is not limited to 60 GHz or varactor diodes and may employ alternative tunable elements (e.g., liquid crystals), this assessment may serve as a reference for future implementations.  

\begin{figure}[t!]
  \centering
    \includegraphics[width=\linewidth]{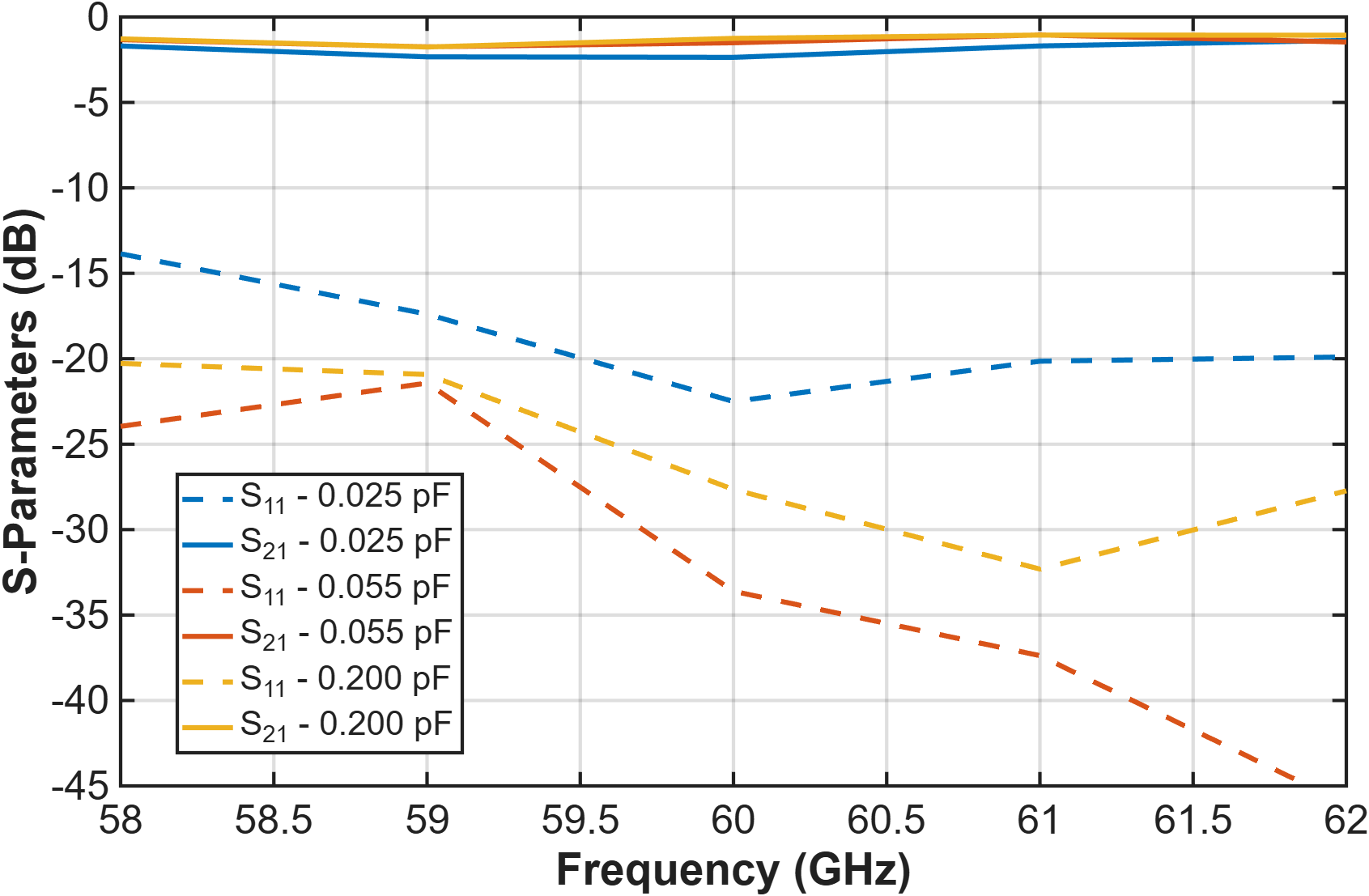}
  \caption{S parameters for a single radiation point for different varactor capacitance values.}
  \label{SparametersElement}
\end{figure}

\begin{figure}[t!]
  \centering
    \includegraphics[width=\linewidth]{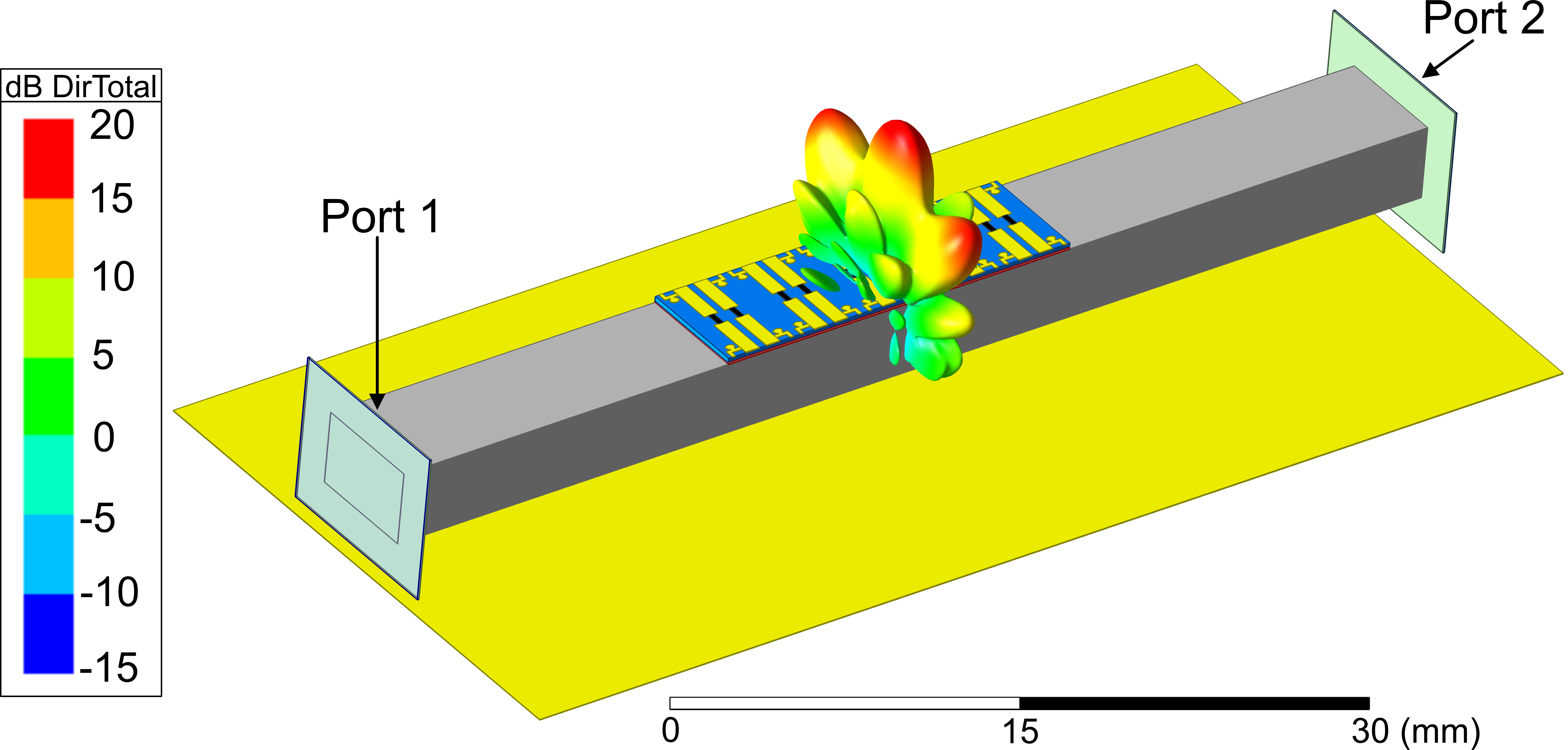}
  \caption{3D polar plot of directivity overlapped with E-pinching antenna.}
  \label{3Dpolar}
\end{figure}

The scattering parameters of the designed antenna are shown in Fig. \ref{SparametersElement}. In these simulations, the waveguide's length is 60 mm (12 $\lambda$ at 60 GHz). When the varactor capacitance is high, the module does not leak energy; consequently, we observe minimal reflection and most of the signal reaches the second port. As the capacitance decreases, more energy leaks from the waveguide, thereby reducing transmission. Based on these figures, one can argue that the radiation zone can be enabled or disabled by adjusting the DC bias applied to the varactor diode. Table \ref{Table1} summarizes the antenna parameters for different capacitance values (assuming a 1 W source power). The OFF state is achieved by using the varactor diode at a high capacitance (0.1 pF). The radiation increases inversely with capacitance, reaching its maximum at 0.025 pF, with 39.6\% of the 1-W input power radiated by the module, a peak system gain of 18.117 (12.58 dB), a peak directivity of 45.747 (16.6 dB), and a side lobe level (SLL) of -9.18 dB. The radiation pattern of this module is shown in Fig. \ref{3Dpolar}. The antenna maintains a half-power beamwidth (HPBW) of 12 degrees across all capacitance values. The main beam is tilted by approximately 5 degrees from broadside, as expected for traveling-wave antennas.

\begin{figure}[t!]
  \centering
    \includegraphics[width=\linewidth]{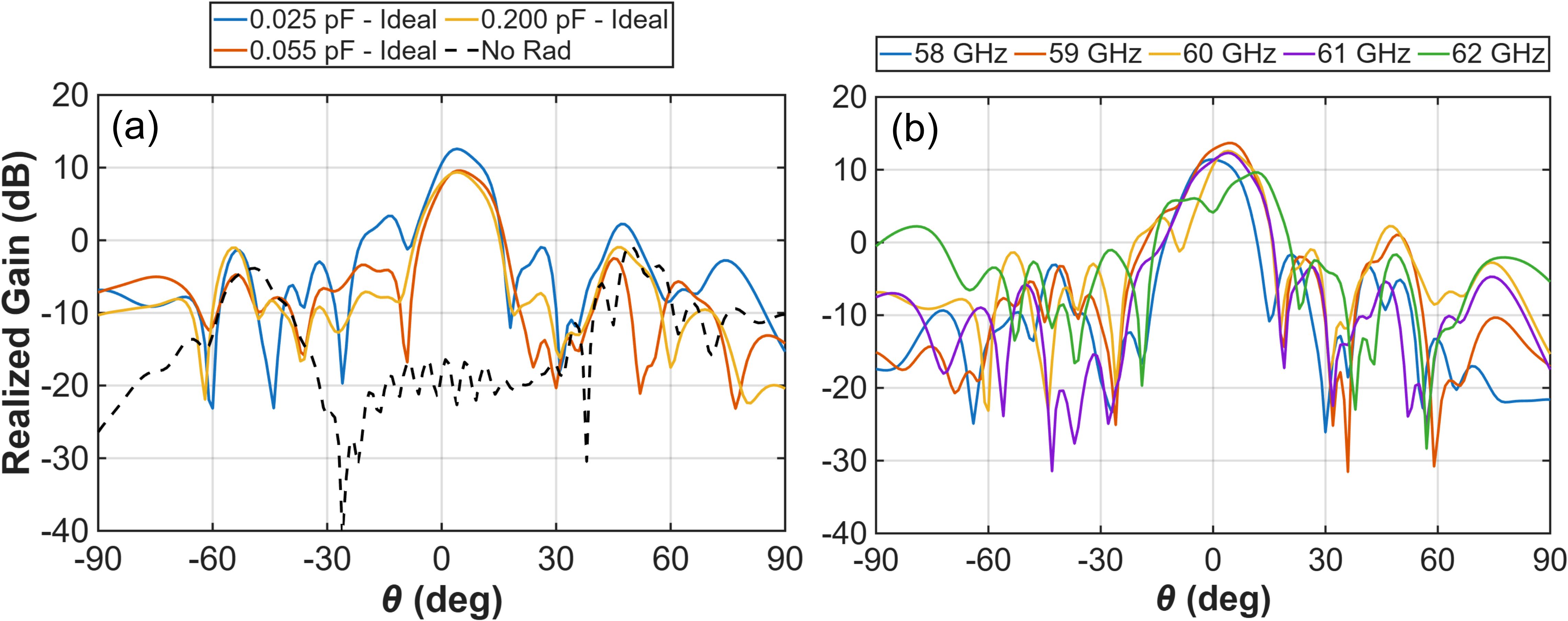} \\[\abovecaptionskip]
  \caption {Total realized gain over the E-plane of the E-pinching antenna for (a) different capacitance values at 60 GHz and (b) different frequencies for 0.025 pF capacitance.}
  \label{3Dpolar-Eplane}
\end{figure}

The realized gain over the E-plane of the proposed antenna at different capacitance levels at 60 GHz is shown in Fig. \ref{3Dpolar-Eplane}(a). We observe that varying the capacitance primarily changes the antenna behavior from non-radiating to radiating. For comparison, we also show the case with no module (waveguide only). Comparing this case with those including the radiation module, we can conclude that the lobes observed around $\pm 60 ^\circ$ are due to edge diffraction. It is also worth emphasizing that we can reduce the overall leakage from this point to a level closer to a true OFF state by optimizing the geometry of metallic strips. For this proof-of-concept demonstration, five metallic strips were used while leaving the optimization for future work.

Fig. \ref{3Dpolar-Eplane}(b) shows the realized gain for a 0.025 pF capacitance as a function of frequency. Although changing the frequency slightly shifts the main beam, it generally remains within the same angular region. This consistent performance is useful for communication applications where the signal level should remain relatively similar across the operating band.

\section{Wireless Communication Simulation}

\begin{figure}[t!]
\centering
\includegraphics[width=\linewidth]{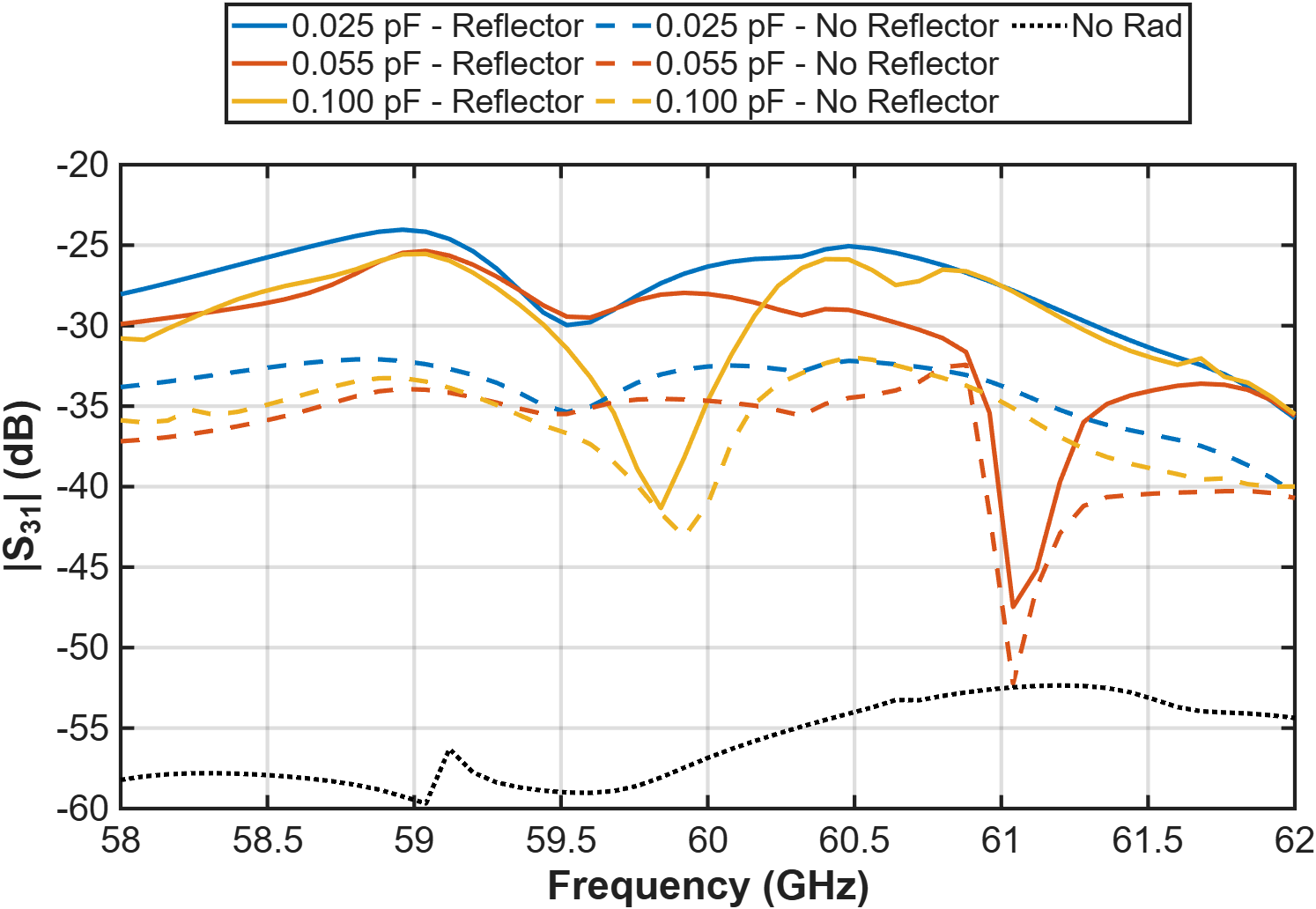}
\caption{Transmission between E-pinching antenna and a patch antenna with and without reflector.}
\label{fig_5}
\end{figure}

We begin by examining the performance of the E-pinching antenna in a wireless communication setting by placing a patch antenna in the LoS of the radiation module to emulate a single user. The E-pinching antenna has a total length of 60 mm and a single radiation module. The patch antenna is placed at 5 cm (10 $\lambda$) from the waveguide and 9 mm from the center of the radiation module to account for the 5-degree tilt of the main beam. The patch antenna ($2\times 1.6$ mm) is designed to resonate at 60 GHz and is implemented on an RT/Duroid 5880 substrate of 4 mm by 4.9 mm (the design is not discussed for brevity). Figure \ref{fig_5} shows the transmission coefficient ($S_{31}$) between the E-pinching antenna (with and without a reflector) and the patch antenna. The modular E-pinching antenna, without a reflector, exhibits transmission parameters adjustable from -41.07 dB to -34.65 dB, yielding a total range of 6.42 dB between the OFF and ON states at 60 GHz. Upon adding the copper reflector, the total range is 8.34 dB, with transmission adjustable from -34.66 dB to -26.32 dB.

\begin{table}[t!]
\setstretch{1.5}
\caption{Radiation parameters of the proposed antenna at 60 GHz for different capacitance values.}
\label{Table1}
\begin{adjustbox}{width=0.49\textwidth}
\begin{tabular}{cccccc}
\hline
Diode capacitance (pF)   & 0.1    & 0.065     & 0.055 & 0.045  & 0.025 \\ \hline
Peak System Gain (dB)   & 3.7 & 9.78  & 9.59 & 10.45 & 12.58 \\
Peak Directivity (dB)   & 12.42 & 16.56 & 15.37 & 14.77 & 16.6 \\
Accepted power (mW)      & 180.57 & 240.05   & 293.02 & 409.77  & 413.88 \\
Radiated power (mW)      & 134.4  & 209.97    & 264.13 & 369.73  & 396.02 \\
Side Lobe Level (dB)       & -3.23  & -10.4  & -12.23 & -10 & -9.18 \\ \hline
\end{tabular}
\end{adjustbox}
\end{table}

Next, the E-pinching antenna is evaluated in the multi-user setup shown in Fig. \ref{fig_2}. This configuration is a small-scale version of a real-world scenario in which the E-pinching antenna is used to communicate with users under both LoS and NLoS conditions. The simulation domain was scaled down to manage computational complexity while preserving the key physical features of the scenario. Even with this reduced domain, the full-wave simulation approaches the limits of our available computational resources and may therefore be subject to some numerical error. Nevertheless, the results demonstrate the expected trends between the ON and OFF states. As shown in Fig. \ref{fig_2}, the second user is located behind a metallic partition and requires a second E-pinching module to receive signals. The total length of the E-pinching antenna system is $l=20$ cm (40 $\lambda$). The conductive blocking wall is positioned at the center of the structure. Each user is equipped with a patch antenna designed for 60 GHz placed at a distance of $10~\lambda$ from the dielectric waveguide and 9 mm from the center of the radiation module. The distance from the first port to the edge of the first radiation module is $l3=4$ cm. An equal spacing is used between the edge of the second radiation module and the second port ($l4 = l3$). Each radiation module has a size of $m=2$ cm, and the edge-to-edge distance between the two modules is $l2=8$ cm. 

\begin{figure}[t!]
\centering
\includegraphics[width=\linewidth]{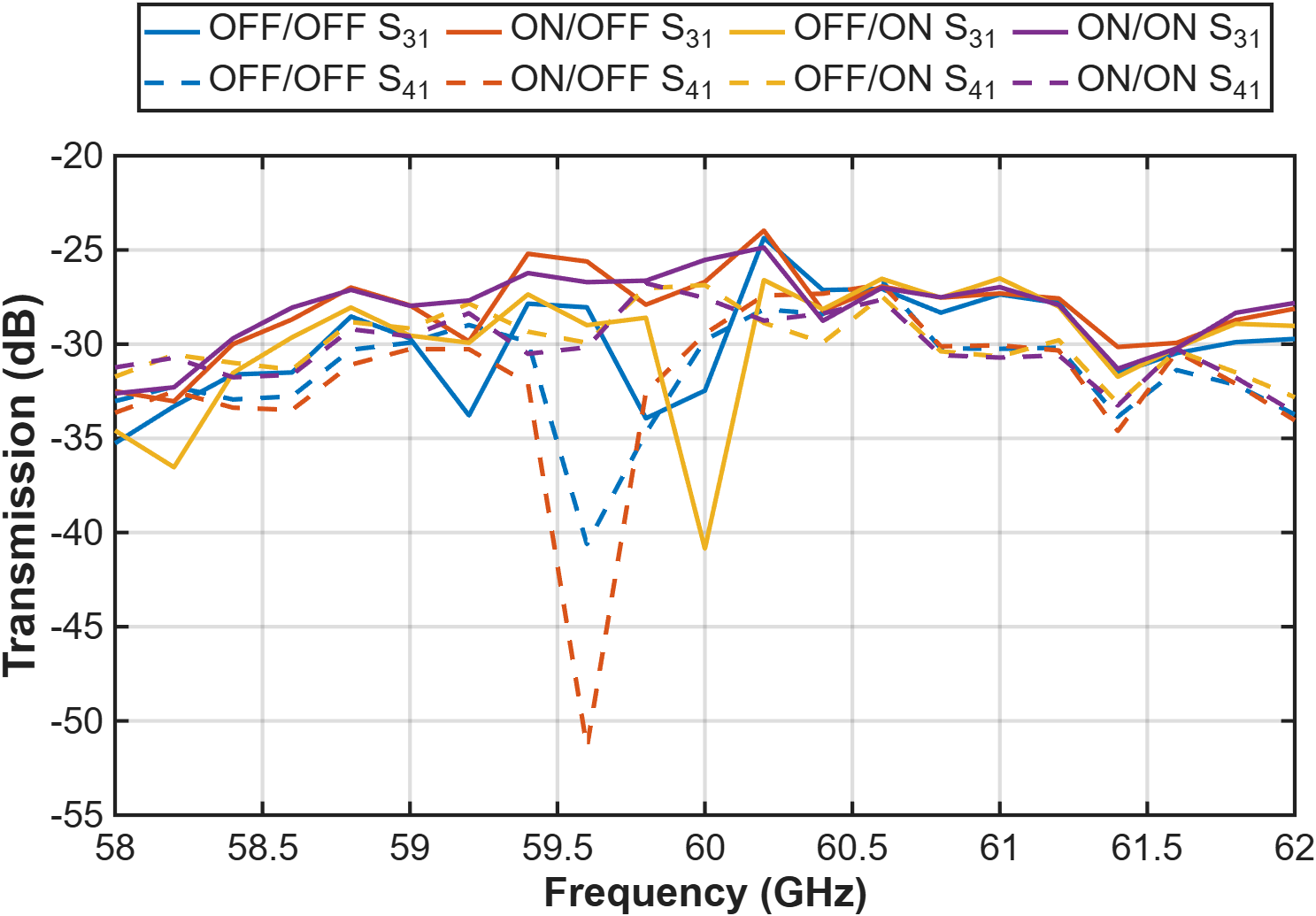}
\caption{Transmission coefficient from the E-pinching antenna with two radiation points and user 1 ($S_{31}$) and user 2 ($S_{41}$).}
\label{fig_7}
\end{figure}

First, we simulate the case when both radiation points are in the OFF state (i.e., the capacitance of all diodes is set to their maximum value ($C = 0.1$ pF)) to set the baseline. Next, we examine the case in which one of the radiation points is turned ON by setting its diode capacitance to its minimum value (0.025 pF). The other module is kept in the OFF state. We observe that, in this case, the E-pinching antenna can provide a reliable signal to both LoS and NLoS users, even in the presence of blockages. Lastly, both modules are configured with the same capacitance value of 0.025 pF, corresponding to the minimum capacitance and maximum radiation from each module. It is worth noting that the LoS user achieves a higher link than the blocked one. This occurs because the second antenna is farther away, accumulating additional dielectric losses and leakage from the first radiation module. Despite these losses, the resulting link quality differs by only approximately 2 dB between the two users, as shown in Fig. \ref{fig_7}. This demonstrates that comparable link performance can be achieved at multiple radiation points using the proposed module configurations. For easier comparison, the mean transmission coefficients in all cases for a frequency band of 1 GHz around 60 GHz are summarized in Table \ref{Table2}. 

The transmission coefficient along the waveguide ($S_{21}$) also changes due to the presence of the radiation modules. Without radiation points, the simulated value is $S_{21} = -1.35$ dB. When two radiation modules are placed on top of the waveguide in the OFF state, the transmission decreases to $S_{21} = -2.36$ dB. Despite this additional attenuation, the loss remains significantly lower than the free-space path loss of 54.03 dB over the same distance. It has also been theoretically demonstrated in \cite{AttenuationMatter?} that pinching antenna systems, even when accounting for waveguide attenuation and propagation losses, can outperform conventional fixed-antenna systems, particularly in NLoS environments and at higher frequencies. 

\begin{table}[t!]
\setstretch{1.5}
\caption{Average Transmission coefficient from 59.5 to 60.5 GHz.}
\label{Table2}
 \begin{adjustbox}{width=0.49\textwidth}
\begin{tabular}{cccccc}
\hline
Module State       & \multicolumn{1}{l}{No module} & OFF/OFF  & ON/OFF  & OFF/ON  & ON/ON   \\ \hline
$S_{31}$ (LoS user)  & -49.650                          & -29.19 & -26.48 & -30.64 & -26.50\\
$S_{41}$ (NLoS User) & -51.667                          & -32.33  & -33.63 & -28.52 & -28.32 \\ \hline
\end{tabular}
\end{adjustbox}
\end{table}

\section{Conclusion and Discussion}

This paper presented the design and operation of an electronically reconfigurable pinching antenna (E-pinching antenna) for controllable mmWave communications. Through full-wave simulations, the proposed architecture was shown to enable electronically controllable radiation points at selected locations along a dielectric waveguide, allowing the radiated signal to be delivered closer to intended users and reducing the dependence on long free-space propagation paths. The radiation pattern of each module can be configured to provide directional radiation suitable for wall- and ceiling-mounted indoor deployments. The results further demonstrated that the radiated power and resulting link strength can be controlled by tuning the switchable component. The potential of the E-pinching antenna as a data highway serving both LoS and NLoS users was also demonstrated. Unlike conventional pinching antennas that rely on mechanical perturbations, the proposed architecture enables electronic control of the radiation points, providing a path toward scalable and reconfigurable pinching-antenna systems.

Several directions can further extend the proposed architecture. The operating frequency can be adapted by modifying the dimensions of the dielectric waveguide and radiation module, while optimization of the metallic strips, module geometry, and switching elements can further improve the electromagnetic response. Alternative switching and tuning technologies may also be investigated to reduce parasitic effects and enable operation at higher frequencies. Furthermore, temporal control of the biasing signals could enable space-time modulation and other advanced reconfiguration strategies. Similarly, jointly considering the electromagnetic behavior of the modules, their placement along the waveguide, and wireless-network deployment requirements represents an important direction toward practical E-pinching antenna systems.

Future work will focus on a comprehensive experimental characterization of the electronically controlled pinching antenna. In particular, measurements will be conducted to quantify the losses introduced by the electronic components and determine their impact on transmission, coupling, and radiation performance. The experimental investigation will compare different operating states and module configurations, evaluate the repeatability of the electronic switching mechanism, and characterize the resulting changes in the radiated fields. Measurements with multiple electronically controlled modules integrated along the same dielectric waveguide will also be conducted to experimentally evaluate independent radiation-point activation, interactions between modules, and reconfigurable radiation-point selection. These experiments will provide the foundation for evaluating more complex electronically controlled pinching-antenna configurations and their integration into practical communication systems.







\bibliographystyle{IEEEtran}

\bibliography{References}

@INPROCEEDINGS{WGSlabAntenna,
  author={Dey, Utpal and Hesselbarth, Jan},
  booktitle={12th European Conference on Antennas and Propagation (EuCAP 2018)}, 
  title={Dielectric slab waveguide based millimeter-wave leaky-wave antennas}, 
  year={2018},
  volume={},
  number={},
  pages={1-5},
  doi={10.1049/cp.2018.0435}}

@INPROCEEDINGS{WGRodAntenna,
  author={Dey, Utpal and Tonn, Julian and Hesselbarth, Jan},
  booktitle={2019 12th German Microwave Conference (GeMiC)}, 
  title={Millimeter-Wave Leaky-Wave Antennas Based on Polymer Rod with Periodic Annular Metal Strips}, 
  year={2019},
  volume={},
  number={},
  pages={9-12},
  doi={10.23919/GEMIC.2019.8698170}}

@article{DOCOMO,
  title={Pinching antenna: Using a dielectric waveguide as an antenna},
  author={Suzuki, Hiroshi Okazaki Yasunori and Kawai, Kunihiro},
  journal={NTT DOCOMO Technical J},
  volume={23},
  number={3},
  pages={5--12},
  year={2022}
}

@ARTICLE{RIS,
  author={Liu, Yuanwei and Liu, Xiao and Mu, Xidong and Hou, Tianwei and Xu, Jiaqi and Di Renzo, Marco and Al-Dhahir, Naofal},
  journal={IEEE Communications Surveys \& Tutorials}, 
  title={Reconfigurable Intelligent Surfaces: Principles and Opportunities}, 
  year={2021},
  volume={23},
  number={3},
  pages={1546-1577},
  doi={10.1109/COMST.2021.3077737}}

@misc{IMS,
      title={Emerging Technologies in Intelligent Metasurfaces: Shaping the Future of Wireless Communications}, 
      author={Jiancheng An and Mérouane Debbah and Tie Jun Cui and Zhi Ning Chen and Chau Yuen},
      year={2025},
      eprint={2411.19754},
      archivePrefix={arXiv},
      primaryClass={cs.IT},
      url={https://arxiv.org/abs/2411.19754}, 
}

@ARTICLE{Fluid,
  author={Wong, Kai-Kit and Shojaeifard, Arman and Tong, Kin-Fai and Zhang, Yangyang},
  journal={IEEE Transactions on Wireless Communications}, 
  title={Fluid Antenna Systems}, 
  year={2021},
  volume={20},
  number={3},
  pages={1950-1962},
  doi={10.1109/TWC.2020.3037595}}

@misc{AttenuationMatter?,
      title={Pinching-Antenna System Design with LoS Blockage: Does In-Waveguide Attenuation Matter?}, 
      author={Yanqing Xu and Zhiguo Ding and Octavia A. Dobre and Tsung-Hui Chang},
      year={2025},
      eprint={2508.07131},
      archivePrefix={arXiv},
      primaryClass={eess.SP},
      url={https://arxiv.org/abs/2508.07131}, 
}

@book{WGBook,
  title={The essence of dielectric waveguides},
  author={Yeh, Cavour and Shimabukuro, Fred I},
  year={2008},
  publisher={Springer}
}

@inbook{BalanisWG,

publisher = {John Wiley \& Sons, Ltd},
isbn = {9781394180042},
title = {Rectangular Cross-Section Waveguides and Cavities},
booktitle = {Balanis' Advanced Engineering Electromagnetics},
chapter = {8},
pages = {365-478},
doi = {https://doi.org/10.1002/9781394180042.ch8},
url = {https://onlinelibrary.wiley.com/doi/abs/10.1002/9781394180042.ch8},
eprint = {https://onlinelibrary.wiley.com/doi/pdf/10.1002/9781394180042.ch8},
year = {2023}
}

@INPROCEEDINGS{RodWGPTFELoss,
  author={Li, Yizhang and Hesselbarth, Jan},
  booktitle={2024 15th German Microwave Conference (GeMiC)}, 
  title={Dual-Polarized Polymer Dielectric Rod Waveguide for 60 GHz Band}, 
  year={2024},
  volume={},
  number={},
  pages={171-174},
  doi={10.23919/GeMiC59120.2024.10485306}}

@ARTICLE{Multipin,
  author={Cao, Lin and Yin, Haifan and Pei, Xilong and Song, Rongguang and Liu, Yuanwei},
  journal={IEEE Wireless Communications Letters}, 
  title={Multi-Pin and Movable-Pin Enabled Discretely-Activated Pinching Antenna Systems}, 
  year={2026},
  volume={15},
  number={},
  pages={590-594},
  doi={10.1109/LWC.2025.3632507}}

@ARTICLE{CommunicationSuperhighways,
  author={Wong, Kai-Kit and Tong, Kin-Fai and Chu, Zhiyuan and Zhang, Yangyang},
  journal={IEEE Wireless Communications}, 
  title={A Vision to Smart Radio Environment: Surface Wave Communication Superhighways}, 
  year={2021},
  volume={28},
  number={1},
  pages={112-119},
  doi={10.1109/MWC.001.2000162}}

@article{shlezinger2021dynamic,
  title={Dynamic metasurface antennas for 6G extreme massive MIMO communications},
  author={Shlezinger, Nir and Alexandropoulos, George C and Imani, Mohammadreza F and Eldar, Yonina C and Smith, David R},
  journal={IEEE Wireless Communications},
  volume={28},
  number={2},
  pages={106--113},
  year={2021},
  publisher={IEEE}
}

@article{trichopoulos2022design,
  title={Design and evaluation of reconfigurable intelligent surfaces in real-world environment},
  author={Trichopoulos, Georgios C and Theofanopoulos, Panagiotis and Kashyap, Bharath and Shekhawat, Aditya and Modi, Anuj and Osman, Tawfik and Kumar, Sanjay and Sengar, Anand and Chang, Arkajyoti and Alkhateeb, Ahmed},
  journal={IEEE Open Journal of the Communications Society},
  volume={3},
  pages={462--474},
  year={2022},
  publisher={IEEE}
}

@article{basar2019wireless,
  title={Wireless communications through reconfigurable intelligent surfaces},
  author={Basar, Ertugrul and Di Renzo, Marco and De Rosny, Julien and Debbah, Merouane and Alouini, Mohamed-Slim and Zhang, Rui},
  journal={IEEE Access},
  volume={7},
  pages={116753--116773},
  year={2019},
  publisher={IEEE}
}

@article{rappaport2019wireless,
  title={Wireless Communications and Applications Above 100 GHz: Opportunities and challenges for 6G and beyond},
  author={Rappaport, Theodore S and Xing, Yunchou and Kanhere, Ojas and Ju, Shihao and Madanayake, Arjuna and Mandal, Soumyajit and Alkhateeb, Ahmed and Trichopoulos, Georgios C},
  journal={IEEE access},
  volume={7},
  pages={78729--78757},
  year={2019},
  publisher={IEEE}
}

@article{bai2014coverage,
  title={Coverage and capacity of millimeter-wave cellular networks},
  author={Bai, Tianyang and Alkhateeb, Ahmed and Heath, Robert W},
  journal={IEEE Communications Magazine},
  volume={52},
  number={9},
  pages={70--77},
  year={2014},
  publisher={IEEE}
}

@inproceedings{kela2016location,
  title={Location based beamforming in 5G ultra-dense networks},
  author={Kela, Petteri and Costa, Mario and Turkka, Jussi and Koivisto, Mike and Werner, Janis and Hakkarainen, Aki and Valkama, Mikko and Jantti, Riku and Leppanen, Kari},
  booktitle={2016 IEEE 84th Vehicular Technology Conference (VTC-Fall)},
  pages={1--7},
  year={2016},
  organization={IEEE}
}

@article{sulyman2016directional,
  title={Directional radio propagation path loss models for millimeter-wave wireless networks in the 28-, 60-, and 73-GHz bands},
  author={Sulyman, Ahmed Iyanda and Alwarafy, Abdulmalik and MacCartney, George R and Rappaport, Theodore S and Alsanie, Abdulhameed},
  journal={IEEE Transactions on Wireless Communications},
  volume={15},
  number={10},
  pages={6939--6947},
  year={2016},
  publisher={IEEE}
}

@INPROCEEDINGS{HE11conversion1,
  author={Pal, Abhijit and Dutta, Debrina and Schneider, Martin},
  booktitle={2025 19th European Conference on Antennas and Propagation (EuCAP)}, 
  title={High-Gain Dielectric Lens Antenna Fed by Circular Dielectric Waveguide for Applications in D-Band}, 
  year={2025},
  volume={},
  number={},
  pages={1-5},
  doi={10.23919/EuCAP63536.2025.10999273}}

@article{HE11conversion2,
  title={A TE01-HE11 mode conversion system},
  author={Niu, Xin-Jian and Wang, Li-Xuan and Liu, Ying-Hui and Wang, Hui and Li, Hong-fu},
  journal={Journal of Infrared, Millimeter, and Terahertz Waves},
  volume={36},
  number={11},
  pages={1101--1111},
  year={2015},
  publisher={Springer}
}

@article{HE11conversion3,
author = {Mun Seok Choe and Kwang Hoon Kim and EunMi Choi},
title = {A comprehensive analysis of a TE11 to HE11 mode converter for an oversized F-band corrugated waveguide},
journal = {Journal of Electromagnetic Waves and Applications},
volume = {27},
number = {17},
pages = {2221--2238},
year = {2013},
publisher = {Taylor \& Francis},
doi = {10.1080/09205071.2013.838147}
}

@ARTICLE{FASPinching,
  author={Ding, Zhiguo and Schober, Robert and Vincent Poor, H.},
  journal={IEEE Transactions on Communications}, 
  title={Flexible-Antenna Systems: A Pinching-Antenna Perspective}, 
  year={2025},
  volume={73},
  number={10},
  pages={9236-9253},
  doi={10.1109/TCOMM.2025.3555866}}

@article{saad2019vision,
  title={A vision of 6G Wireless Systems: Applications, trends, technologies, and open research problems},
  author={Saad, Walid and Bennis, Mehdi and Chen, Mingzhe},
  journal={IEEE network},
  volume={34},
  number={3},
  pages={134--142},
  year={2019},
  publisher={IEEE}
}

@misc{1220566,
  author = {Nada T. Golmie and Marc Leh and Miller Higgins},
  title = {NextG Communications Research and Development Gaps Report},
  year = {2023},
  month = {2023-11-15 05:11:00},
  publisher = {Special Publication (NIST SP), National Institute of Standards and Technology, Gaithersburg, MD},
  url = {https://tsapps.nist.gov/publication/get_pdf.cfm?pub_id=956201},
  doi = {https://doi.org/10.6028/NIST.SP.1293},
  language = {en},
}

@ARTICLE{Kiat,
  author={New, Wee Kiat and Wong, Kai-Kit and Wang, Chao and Chae, Chan-Byoung and Murch, Ross and Jafarkhani, Hamid and Hao, Yang},
  journal={IEEE Journal on Selected Areas in Communications}, 
  title={Fluid Antenna Systems: Redefining Reconfigurable Wireless Communications}, 
  year={2026},
  volume={44},
  number={},
  pages={1013-1044},
  doi={10.1109/JSAC.2025.3632097}}

@misc{zheng2025reconfigurableantennas6gtechnologies,
      title={Reconfigurable Antennas for 6G: Technologies, Prototypes, Architectures, and Signal Processing}, 
      author={Pinjun Zheng and Ruiqi Wang and Yuchen Zhang and Md. Jahangir Hossain and Anas Chaaban and Atif Shamim and Tareq Y. Al-Naffouri},
      year={2025},
      eprint={2506.00657},
      archivePrefix={arXiv},
      primaryClass={eess.SP},
      url={https://arxiv.org/abs/2506.00657}, 
}

@ARTICLE{LWAsHum,
  author={Suntives, Asanee and Hum, Sean V.},
  journal={IEEE Transactions on Antennas and Propagation}, 
  title={A Fixed-Frequency Beam-Steerable Half-Mode Substrate Integrated Waveguide Leaky-Wave Antenna}, 
  year={2012},
  volume={60},
  number={5},
  pages={2540-2544},
  doi={10.1109/TAP.2012.2189726}}

@INPROCEEDINGS{WGConversion4,
  author={Trinh, T.N. and Malherbe, J.A.G. and Mittra, R.},
  booktitle={1980 IEEE MTT-S International Microwave symposium Digest}, 
  title={A Metal-to-Dielectric Waveguide Transition with Application to Millimeter-Wave Integrated Circuits}, 
  year={1980},
  volume={},
  number={},
  pages={205-207},
  doi={10.1109/MWSYM.1980.1124234}}

@techreport{WGConversion5,
  title={Investigation of Metal to Dielectric Waveguide Transition and the Coupling Characteristics of Dielectric Waveguides of Rectangular Cross Section.},
  author={Trinh, Trang Nha},
  year={1980}
}

@article{Yang2025Pinching,
  author  = {Zheng Yang and Ning Wang and Yanshi Sun and
             Zhiguo Ding and Robert Schober and
             George K. Karagiannidis and
             Vincent W. S. Wong and
             Octavia A. Dobre},
  title   = {Pinching Antennas: Principles, Applications and Challenges},
  journal = {arXiv preprint arXiv:2501.10753},
  year    = {2025},
  doi      = {10.48550/arXiv.2501.10753}
}

@INPROCEEDINGS{Epinch,
  author={Plata-Orozco, Fernando and Imani, Mohammadreza F.},
  booktitle={2026 IEEE International Symposium on Antennas and Propagation and USNC-URSI Radio Science Meeting (AP-S/USNC-URSI)}, 
  title={Modular Electronically Reconfigurable Pinching Antenna}, 
  year={2026},
  volume={},
  number={},
  pages={1531-1534},
  doi={10.1109/AP-S/USNC-URSI60190.2026.11675037}}

\vfill

\end{document}